\documentclass[pra,twocolumn,superscriptaddress]{revtex4}
\usepackage{amsmath}
\usepackage{amstext}
\usepackage{latexsym}
\usepackage[dvips]{graphicx}

\newcommand{\ket}[1]{\left\vert#1\right\rangle}
\newcommand{\bra}[1]{\left\langle#1\right\vert}

\newcommand{\beq}{\begin{equation}}
\newcommand{\eeq}{\end{equation}}

\begin{document}
\title{Fractal Fidelity as a signature of Quantum Chaos}

\author{Franco Pellegrini}
\affiliation{NEST-CNR-INFM \& Scuola Normale Superiore, Piazza dei
	Cavalieri 7, I-56126 Pisa, Italy}

\author{Simone Montangero}
\affiliation{NEST-CNR-INFM \& Scuola Normale Superiore, Piazza dei
	Cavalieri 7, I-56126 Pisa, Italy}
\affiliation
{Institut f\"ur Theoretische Festk\"orperphysik and DFG-Center for 
Functional Nanostructures (CFN),
Universit\"at Karlsruhe, D-76128 Karlsruhe, Germany.}

\date{\today}
\begin{abstract}
We analyze the fidelity of a quantum
simulation and we show that it
displays fractal fluctuations iff the simulated dynamics is chaotic.
This analysis allows us to investigate a given simulated dynamics 
without any prior knowledge.
In the case of integrable dynamics, the appearance
of fidelity fractal fluctuations is a signal of a highly corrupted simulation. 
We conjecture that fidelity fractal fluctuations are a signature of
the appearance of quantum chaos.  Our analysis can be realized already
by a few qubit quantum processor. 
\end{abstract}

\maketitle
Quantum signature of classical chaotic systems have been investigated widely 
in the last decades~\cite{haake}. Within the development of quantum
information  theory,
a new interest on quantum chaotic systems arose motivated by the fact
that they are interesting testbeds for quantum simulations.
Indeed, the quantum simulation of quantum chaotic systems is possible
and some  efficient 
algorithms, with respect to known classical ones, have been
found~\cite{monta01,baker,kickrot}. 
Moreover, it has been shown that classical simulations
of quantum chaotic systems are a difficult problem for classical
computers due to the 
high presence of entanglement in the fully chaotic
regime~\cite{monta04, prosen}.
In a more general context the fidelity, the response to a Hamiltonian perturbation of a 
quantum system, has attracted a lot of attention since its introduction
by Peres~\cite{peres}. Indeed it has been shown that the fidelity is of fundamental importance 
for the understanding of a system dynamics as
it has been used to characterize the system integrability~\cite{emerson}, 
the feasibility of a quantum simulation~\cite{monta01,monta02} and of quantum 
communication protocols~\cite{dechiara, romito, bose}, 
the quantum-classical transition~\cite{casati0}, 
the signature of the chaos border~\cite{tsallis}, the effects of a bath~\cite{zurek1} 
the environment induced decoherence~\cite{zurek2} and the characterization of 
quantum phase transitions~\cite{zanardi, rossini,cucchietti}. It has
also been shown that, under given conditions, the fidelity 
recall classical properties of chaotic systems as its decay rate is
given by the classical Lyapunov exponent~\cite{jacquod,pastawski}. 

Differently from mentioned previous studies on the variance of
fidelity fluctuations that characterize the system
response (averaged over different initial conditions or an ensemble of different
perturbations) starting from its complete knowledge \cite{varfluc}, we approach the
problem as for classical complex systems signal analysis. Suppose you have a black box 
you do not have a complete control on as, for example, a quantum computer with some given but
unknown imperfections. Any system outcome, i.e. a result of a
measurement, might be corrupted: in our example, the result of a quantum
computation in presence of hardware imperfections. What might be learnt from such signal
without completely characterize the system and/or without affording
the cost of repetitions of the computation?
In this scenario, we focus on the problem of finding a clear signature of chaos in 
a quantum systems: it is well known that in classical systems with mixed phase space, 
a fractal region arises at the border between integrable islands and the chaotic
sea~\cite{zaslavsky}. These regions are known to be responsible for fractal conductance 
fluctuations  as a function of a given parameter (Aharonov-Bohm flux) 
in open quantum systems in the semiclassical regime~\cite{blumel} and
it has been shown that fractal fluctuations survive also in the deep quantum
regime~\cite{casati}. Moreover, fractal properties of systems wave functions have been 
shown to appear under given conditions~\cite{berry}. 
Here we show that in a unitary evolution, in particular
in a quantum computer running a quantum simulation in the presence
of static imperfections, fidelity fractal fluctuations naturally arise 
as a function of time. 
We demonstrate that the fractal dimension of such fluctuations strikingly 
depends on the chaoticity of the dynamics: For integrable dynamics the fidelity 
dimension is integer while for chaotic dynamics it is fractional. 
This sensitivity is not restricted to a fully chaotic phase space:
fidelity fractal fluctuations arise also in the chaotic regions of a mixed phase
space.  We establish this connection and we present two possible
applications of this analisys. First of all it can be used as a testbed
for a given quantum computer hardware: If running a simple
(integrable) algorithm fractal fidelity fluctuations appears the
hardware is not realible as quantum chaos has set in \cite{monta01}. 
Secondly,  we present a 
method to investigate an unknown phase space (Husimi or Wigner function) of a
general quantum system. Indeed the fractal dimension of the fidelity
of a given quantum system can be used to extract the
presence of a chaotic region in the system phase space. This is of fundamental
importance as the simulation of quantum system is one of the most
general application of quantum computation, however it is severely
limited by the lack of methods, different from the complete wave
function tomography, to extract information from the final wave function~\cite{feynman,casmon}. 
In~\cite{emerson}, it has been shown that the average fidelity can be 
measured in a quantum processor;
the sawtooth map algorithm has been recently performed on a three 
qubit NMR quantum processor~\cite{cory1} and the fidelity
experimentally measured on a three qubits quantum processor~\cite{laflamme}, 
thus, the proposed protocol is at the edge of present day technology. 

This paper is structured as follows: in Section I we present the
testbed for our findings, the Sawtooth Map and its simulation on a
quantum computer in presence of imperfections. 
In Section II we introduce the tools to characterize the fidelity
fractal fluctuations and we analyze the fidelity fractal dimension 
in different scenarios. Finally in Section III we introduce the phase
space tomography via fidelity fractal dimension analysis and Section
IV is devoted to our conclusions.  

\section{Sawtooth Map}
As an example of quantum system with complex dynamics we use the quantum sawtooth map. 
The correspondent classical map is characterized by  a range of very different 
dynamics depending on system parameters varying from completely integrable, 
semi integrable to chaotic dynamics. 
It is defined by the equations
\begin{equation}
{\cal T}:\left\{ 
\begin{array}{ll}  
n_{i+1} &=  n_i + k (\theta_i - \pi) \\
\\
\theta_{i+1} &=  \theta_i + T \: n_{i+1}
\end{array}
\right. ,
\label{csaw}
\end{equation}
where $(n,\theta)$ are the conjugate action-variables $(0 \le \theta
< 2\pi)$, $k$ is the kick strength and $T$ the time between
consecutives kicks. The dynamics is characterized by the parameter $K=kT$: it
is chaotic for $K>0$, $K<-4$, 
integrable for $K= -3,-2,-1,0$ and  with mixed phase space in the 
remaining regions~\cite{cary}. This map belongs to the kicked maps family, 
and it describes a system undergoing free evolution and subject to a kick every time 
$T$. The kick strength is proportional to $\theta^2$.
\begin{figure}[t]
  \begin{center}
    \includegraphics[scale=0.3]{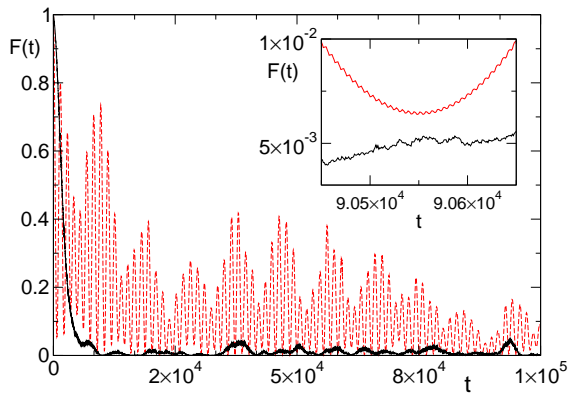}
    \caption{Fidelity as a function of time for integrable $K=-1$ (red, dashed) and chaotic regime
      $K=\sqrt{2}$ (black, full), with $\epsilon= 10^{-4}$ and $n_q=8$. 
      Inset: magnification of the behavior for long times.}
    \label{fidelity}
  \end{center}
\end{figure}
After canonical quantization of the action angle variables, 
the correspondent quantum map is defined 
by the Floquet operator (time evolution operator for a period $T$)
\beq
\hat U = e^{-\frac{i}{2} T \hat n^2} e^{ik (\hat \theta -\pi)^2/2},
\label{qsaw}
\eeq
where $\hat{n}=-i\partial/\partial\theta$ and $\psi(\theta + 2\pi) =
\psi(\theta)$ (we set $\hbar=1$). The quantization rule is
 $[\hat n,\hat \theta]= -i$. 
\begin{figure}[t]
  \begin{center}
\includegraphics[scale=0.3]{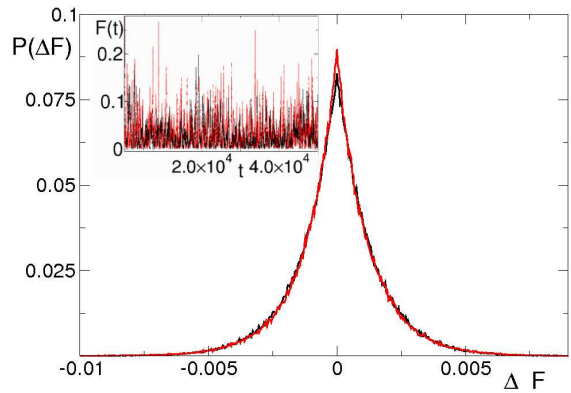}
    \caption{Fidelity (Inset) and distribution $P(\Delta F)$ of fidelity fluctuations
    $\Delta F= F_{i+1}-F_{i}$ for $n_q=6$, $\epsilon= 10^{-3}$ for
    chaotic (black line) and integrable (red line) dynamics.}
    \label{distosc}
  \end{center}
\end{figure}
The map dynamics is still governed by the classical parameter $K$, but a 
key role is played also by $k=K/T$ which rules the quantum-classical transition
($k \to \infty$ is the classical limit, $T=2 \pi/N$ with $N=2^{n_q}$
the Hilbert space size).
In~\cite{monta01} an efficient quantum algorithm has been introduced to simulate
this map and the effects of static imperfections, present in
any experimental apparatus, have been studied~\cite{monta01, monta02}.
The imperfections considered are both residual coupling between qubits and 
fluctuations of the single qubits level spacing. As the effects of the static 
imperfections are almost independent from their exact expression in terms of 
operators if there is a stronger leading dynamics (the quantum algorithm in our case)
here we consider only the latter kind of error. 
The complete Hamiltonian that describes the hardware of the quantum computer is then
\beq
H = \sum_{i=1}^{n_q} (\Delta + \delta_i) \sigma_i^z.
\label{errors}
\eeq
Here $\Delta$ is the mean qubit level spacing, $\delta_i$ are the fluctuations 
of the qubit level spacing taken randomly and uniformly 
in the interval $[-\epsilon; \epsilon]$ (constant in time) and $\sigma_i^z$ are the 
Pauli matrices. 
The Floquet operator in the presence of errors is then defined by the quantum gates 
needed to simulate a period of the map (\ref{qsaw}) with the action of the 
Hamiltonian (\ref{errors}).
The Floquet eigenvalues and eigenvectors statistics have been studied in~\cite{monta02,monta03} and it 
has been shown that eventually with growing imperfections strength the system ends 
to be chaotic regardless the simulated map dynamics~\cite{monta02}. This crossover, is governed
by the imperfections strength $\epsilon$ and it has been shown that in the case of 
integrable dynamics $\epsilon_c \propto n_q^{-5/2}$: this threshold
for chaos appearance has 
been estimated from the breakdown of perturbation theory and confirmed numerically 
\cite{monta03}. 

\section{Fractal fidelity fluctuations}
We study the time fluctuations of the fidelity of a quantum computation.
The fidelity is defined (for pure states) as
\beq
F(t) = |\bra{\psi_\epsilon (t)}| \psi(t)\rangle |^2, 
\eeq
that is the overlap as a function of time of the wave function computed with
the exact time evolution $\ket{\psi(t)}$ and the wave function in the presence
of static imperfections $\ket{\psi_\epsilon (t)}$. The fidelity starts
from one, it decays up to a saturation value and then oscillates around it after a
transient time $t^*$. 
\begin{figure}[t]
  \begin{center}
    \includegraphics[scale=0.28]{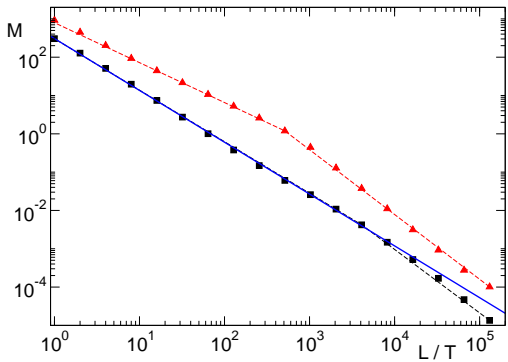}
    \caption{Results of the modified box counting algorithm for the fidelity of
	Fig.\ref{fidelity}: integrable (red triangles) and chaotic
	dynamics (squares). The blue full line is a guide for the eye
	proportional to $L^{1.36}$. Here $t^* \sim 10^4$. 
	The resulting fractal dimensions are $D \sim 1.06$ for the integrable
	case and $D \sim 1.36$ for the chaotic dynamics.}
    \label{fdim}
  \end{center}
\end{figure}

In Fig.~\ref{fidelity} we report two typical fidelities
as a function of time for a chaotic and an integrable dynamics. 
Notice that the signals are not averaged over different static imperfections 
configuration: they could be the direct output of an ``echo
experiment". We stress the fact that the two clear distinct behavior
of the fidelity fluctuations shown in the inset of Fig.~\ref{fidelity}
reflect the presence, in the system classical limit, of a continuum set of typical
frequencies of the system dynamics in the chaotic case differently
from the discrete set of harmonics present in integrable systems. 
Even though in Fig.~\ref{fidelity} the two curves appears
pretty different and one could guess the underlaying behavior of the
system under study, there are case where this is not the case. In Fig.~\ref{distosc}
we show an example where not only the fidelity time dependence but also
the distributions of the fluctuations are not distinguishable. In order
to extract the information on the underlying dynamics one should use
more sophisticated tools such as the study of the fractal dimension of
the signal which detects, as for the classical cases, 
the presence of a ``complex'' set of typical frequencies of
the system in the chaotic case and a ``regular'' spectrum in the
integrable one. 

The fractal dimension of the signal is measured by means of the modified
box counting algorithm~\cite{box1}.
In the standard box counting algorithm the fractal
dimension $D$ of the signal is obtained by covering the data with a
grid of square boxes of size $L^2$. 
\begin{figure}[t]
  \begin{center}
    \includegraphics[scale=0.35]{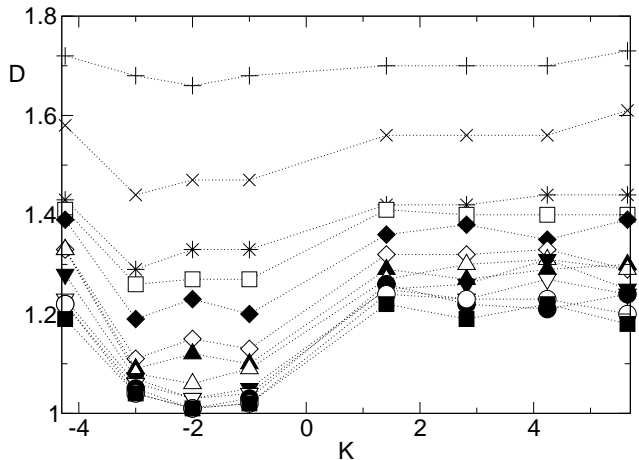}
    \caption{Fractal dimension of the fidelity fluctuations $D$ as a function of
      Sawtooth Map parameter $K$, imperfections strength from
      $\epsilon=10^{-2}$ to $\epsilon=10^{-6}$ (from top to bottom)
      for $n_q=8$  averaged over $N_R < 5$ different imperfections realizations.}
    \label{dk}
  \end{center}
\end{figure}
The number $M(L)$ of boxes
needed to cover the curve is recorded as a function of the box
size $L$. The (fractal) dimension $D$ of the curve is then defined as 
\beq
	D = - \lim_{L \to 0} \log_L M(L). 
\label{fractal}
\eeq
One finds $D=1$ for a straight line, while $D=2$
for a periodic curve. Indeed, for times much larger than the period, a
periodic curve covers uniformly a rectangular region.  Any given value of $D$ in
between of these integer values is a signal of the fractality of the curve. 
The modified algorithm of Ref.~\cite{box1} follows the same lines but 
uses rectangular boxes of size $ L \times \Delta_i$ ($\Delta_i$ is the largest 
excursion  of the curve in the region $L$). Then, the number 
$	M(L) = \frac{\sum_i  \Delta_i}{L} $
is computed. For any curve that lays in a plane, a region of box lengths $L_{min} 
< L < L_{max}$ exists where $M \propto L^D$. Outside this region one
either finds $D=1$ or $D=2$: The first equality ($D=1$) holds for
$L<L_{min}$ and it is due to the coarse grain introduced
by the discrete map time evolution. The second limit ($D=2$) is obtained for
$L>L_{max}$ and it is due to the finite length of the analyzed time
series. The boundaries $L_{min},L_{max}$ have to be chosen properly for
any time series. In our analysis $L_{min}=T$ while $L_{max} \sim
\epsilon^{-\alpha}$, where $\alpha$ increases  
with $n_q$. 
\begin{figure}[t]
  \begin{center}
    \includegraphics[scale=0.33]{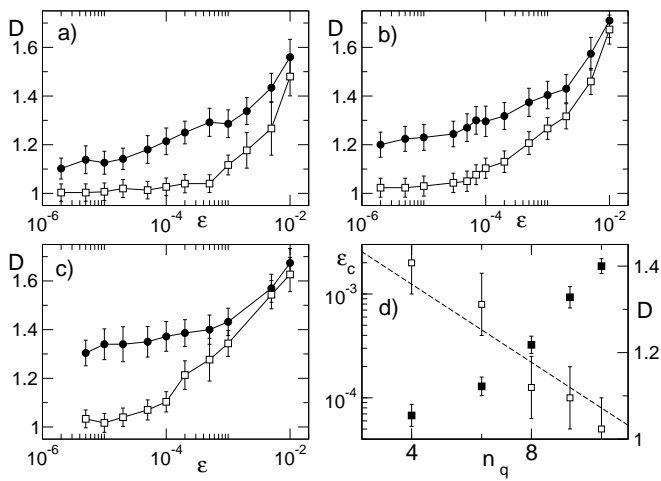}
    \caption{a,b,c) Fidelity fractal dimension $D$ as a function of
    imperfections strength $\epsilon$ averaged over different $K$
    for integrable (white) and chaotic regime (black) for
    $n_q=6,8,10$. \newline 
    d) Scaling of the critical imperfections strength $\epsilon_c \sim n_q^{-2.5}$
     (white squares, left axis) and limiting fractal dimension $D(\epsilon=0)$ as
    a function of $n_q$ (black squares, right axis).}
    \label{deps}
  \end{center}
\end{figure}
In Fig.~\ref{fdim} we report an example of this analysis for the long
time behavior of the two signals of 
Fig.~\ref{fidelity}. We analyzed the time series starting from a given
time $t^*$ to exclude the initial transient decay. 
In Fig.~\ref{dk} we summarize our results reporting the fractal dimension 
of the fidelity for different values of $K$ spanning a wide range 
of different dynamics
(excluding mixed phase space) and for different $\epsilon$ values. Again, 
the signature of chaotic dynamics is striking. Notice that in the chaotic region
the fractal dimension of the fidelity fluctuations appears to be $K$
independent. Thus, we average over different $K$ in both regimes
(integrable and chaotic) to study the dependence of $D$ as a function
of the number of qubits and the imperfections strength.
In Fig.~\ref{deps}~(a, b, c) 
we show the extended analysis of the fractal dimension of the fidelity
fluctuations as a function of the number of qubits in the quantum
hardware and imperfections strength for the 
two distinct dynamics. The difference is striking: for chaotic dynamics 
the dimension is fractal for any value of the imperfections strength. On the contrary, 
for integrable dynamics the dimension of the fidelity fluctuations is one until
a critical value of the imperfections strength $\epsilon_c$. In Fig.~\ref{deps}~d, 
we show the scaling of $\epsilon_c$ as a function of the number of qubits in the system.
Although the scaling is on a very small range of parameter, it scales as the critical 
value for which the integrable dynamics became chaotic due to the
imperfections effect~\cite{monta03}. 
It is then clear, that above $\epsilon_c$ the system is chaotic due to 
the imperfections and regardless of the map dynamics: this transition is reflected by the 
appearance of fractal fluctuations. 
Increasing further the imperfections strength eventually the fidelity
loses all the information regarding the underlying dynamics and
indeed the fractal dimensions related to the two distinct dynamics are
equal. The error bars are due to the sensitivity of the
results to the choice of $L_{min}$ and $L_{max}$. 
In Fig.~\ref{deps}~d
we also show the scaling of the fractal dimension $D(n)$ as a function of the number 
of qubits in the limit $\epsilon \to 0$ for the chaotic regime. The signature of 
quantum chaos is again striking and the difference between the two regimes 
grows with the system size.\\

Notice that this analysis can be performed already
with a few qubits quantum computer (see Fig.~\ref{deps} a). The biggest
obstacle to a possible implementation is the length of the time series 
and the correspondent necessary long coherent times. However, we
checked that already with
boxes of sizes $L \sim 10$ in a time series of few hundred points,
the different scalings in the two cases are clearly visible. 
It is also possible, for small $\epsilon$, to perform the same analysis on the initial
part of the fidelity after subtracting the
average decay  (data not shown). Although this procedure might suffer
from errors due to fitting procedure of the average behavior, it paves
the way to an efficient measurement of the fractal dimension of the
fidelity fluctuations.
\begin{figure}[t]
  \begin{center}
\includegraphics[width=3.2cm]{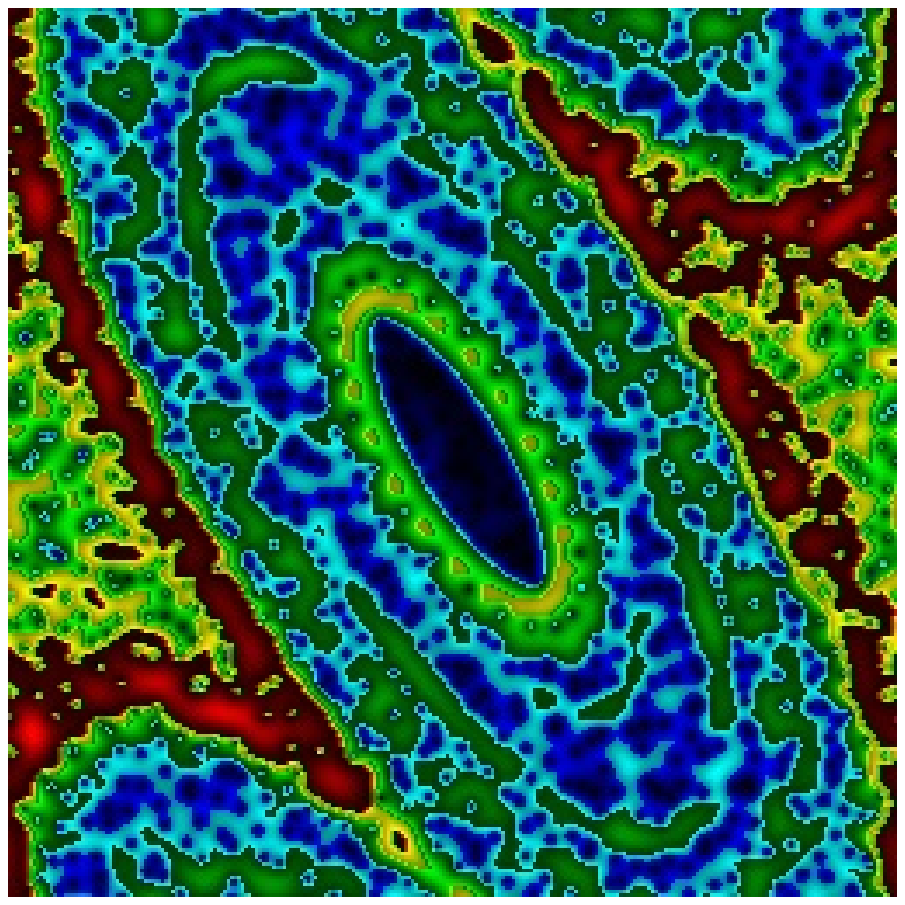}
\includegraphics[width=3.2cm]{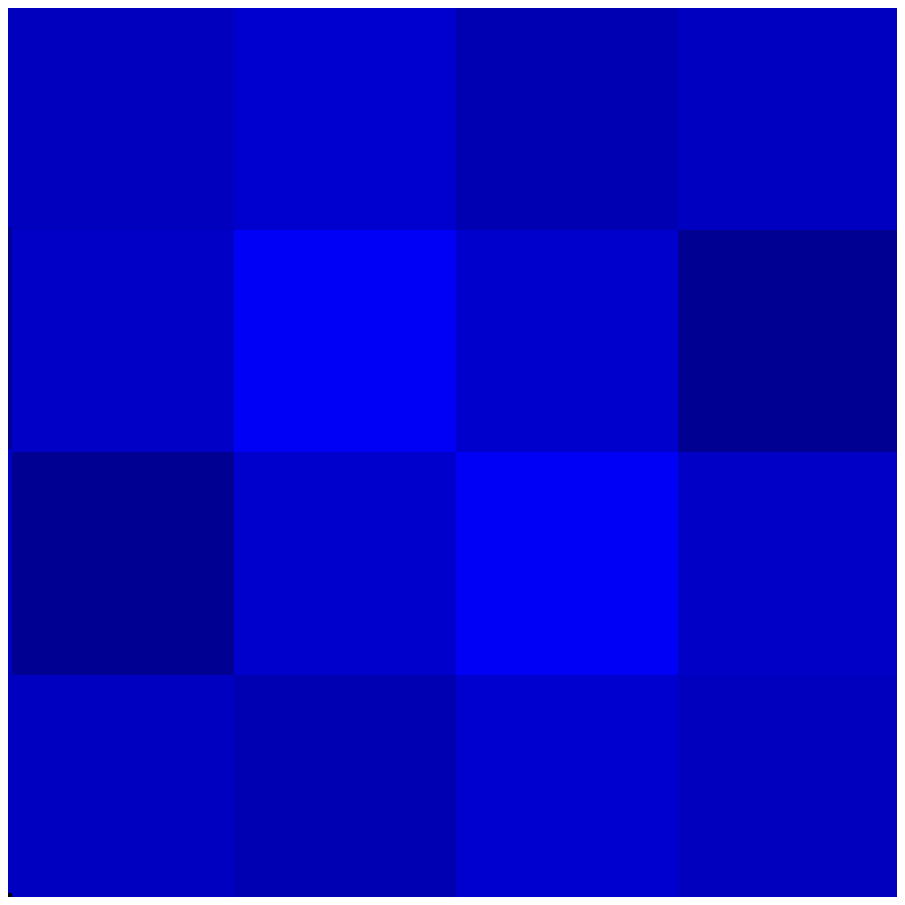}
\includegraphics[width=3.2cm]{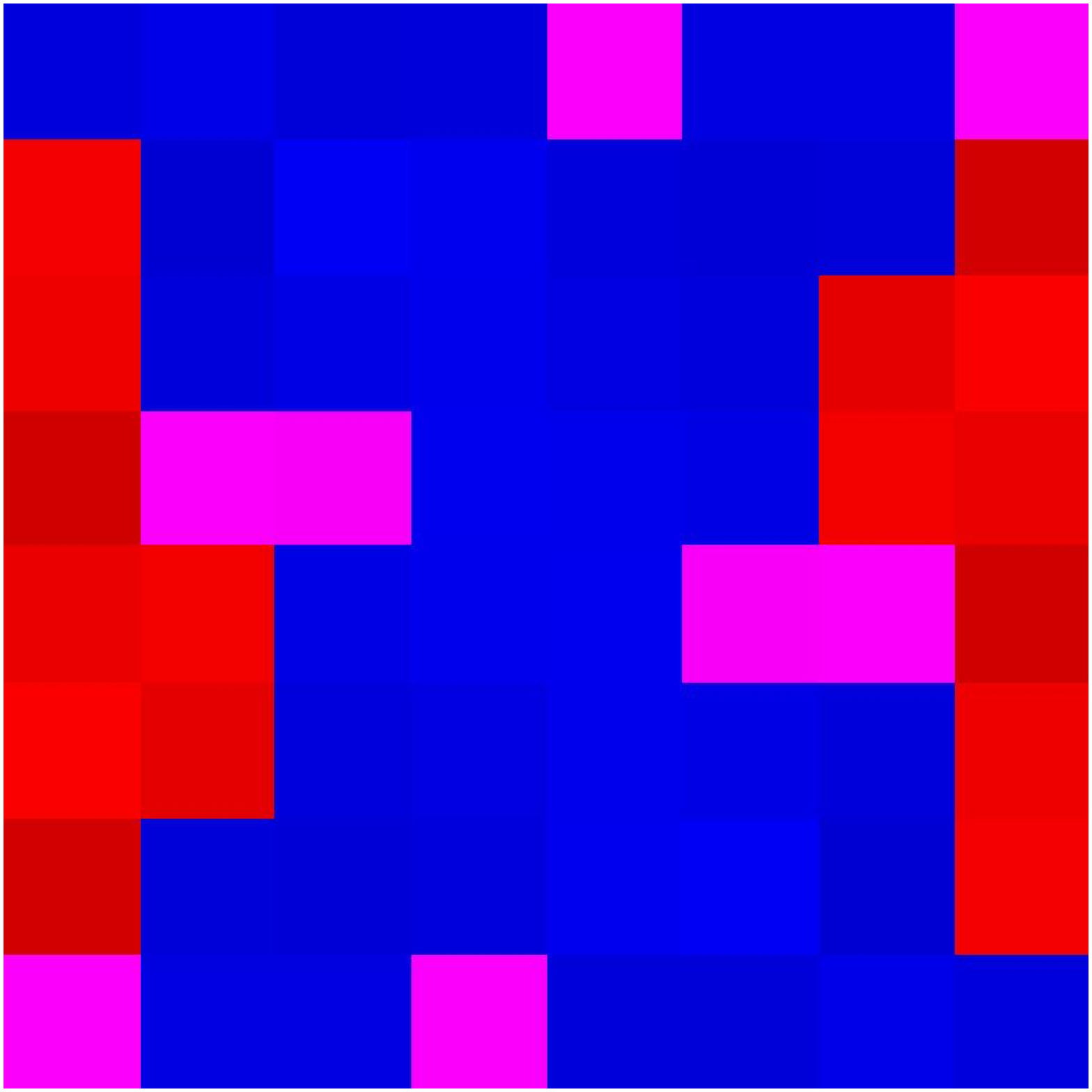}
\includegraphics[width=3.2cm]{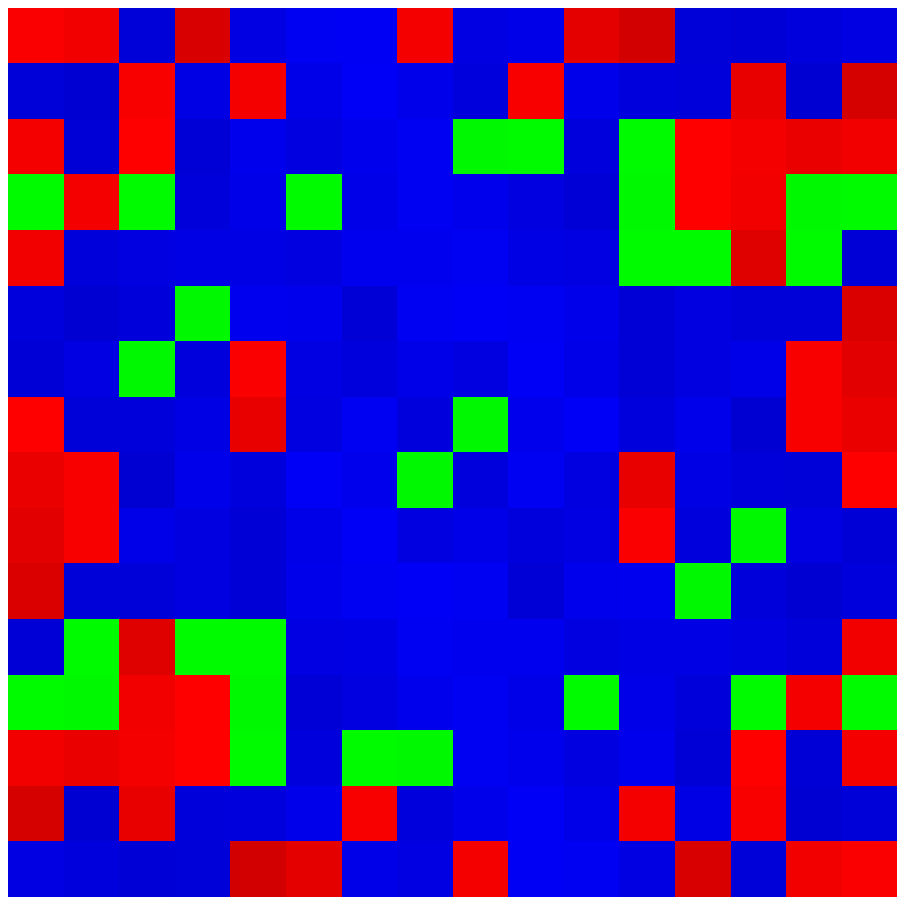}
    \caption{(Color online) Upper left panel: Husimi function of the Sawtooth Map phase space
    $(\theta,n)$ for an
    initial gaussian condition inside
    the chaotic sea ($D=1.34$)
     for $\epsilon=2\cdot 10^{-5}$, $n_q=10$, $t= 2^{14}$. 
     Other panels:  Phase space tomography with different resolutions.
     Fractal dimension of the fidelity fluctuations as a
     function of the initial condition ($4 \times 4, 8 \times 8$ and
     $16 \times 16$) for $n_q=10$ and 
     $\epsilon=2\cdot 10^{-5}$, $N_R=1$.
     The color code goes from blue, yellow to red to represent higher
     values of probability or fractal dimension.}
    \label{icond}
  \end{center}
\end{figure}
\section{Phase space Tomography}
Finally, we concentrate on the fidelity fluctuations as a function of the initial 
condition in the case of mixed phase space. In the upper left panel of
Fig.\ref{icond} we report the Husimi function of
the phase space  of the Sawtooth Map for $K=-2.1$, $T=2 \pi/N$
where the chaotic sea and the integrable islands are clearly 
visible~\cite{husimi,monta01}. The
initial conditions is a gaussian packet which satisfy the minimal
indetermination Heisenberg condition centered 
inside the chaotic sea. 
In the lower panels of Fig.\ref{icond} we show the result of the phase
space tomography with the proposed method with a discretization of
$4 \times 4$, $8 \times 8$ and $16 \times 16$ initial conditions. 
For every initial condition, a gaussian packet centered in
$(n_0,\theta_0)$, we computed the fractal dimension of the fidelity
fluctuations $D(n_0,\theta_0)$ and plotted the resulting contour plot
from $D=1$ (black) to $D=D_{max}$ (red). 
We found again that the dimension of the fidelity depends on the
underlying dynamics: a fractional dimension is found for the chaotic
sea while it is integer when the dynamics lies inside the integrable
islands. The result is a partition of the phase space in chaotic and
integrable regions which can be refined increasing the number
of different initial conditions (notice that the resolution of the
Husimi is of the order of $N^2=2^{20}$ points).
Thus, this analysis can be used as a tool to scan an unknown
phase space of a given Hamiltonian dynamics to discern between
integrable and chaotic phase space regions. \\

\section{Conclusions}
In conclusion, the fidelity time fluctuations are a signature 
of quantum chaotic dynamics. They reflect, as for classically chaotic 
signals, the presence of an ``almost-continuum'' of frequencies in the
system dynamics compared to the ``discrete'' set of typical frequencies of integrable
systems. A similar behavior has been already observed in the context of
quantum communication in spin chains \cite{dechiara}.
We propose this kind of analysis to
investigate  an unknown phase space of a complex quantum
system and to test the effects of static imperfections of quantum
hardware. We stress that the sensitivity of the fidelity fluctuations 
of a simulated system dynamics is valid also in the limit of
imperfections strength going to zero, i.e. 
they will be present in any experimental quantum hardware in case of
chaotic or mixed phase space dynamics.  
Similar results are likely to be found 
in presence of time dependent imperfections (errors) in
the quantum system along the lines of~\cite{facchi}. 

We thank R. Fazio, G. Benenti and D. Frustaglia for useful discussions.
This work has been supported by the ``Quantum Information
Program'' of Centro De Giorgi of Scuola Normale Superiore and by EUROSQIP. 
SM acknowledges support from the Alexander Von  Humboldt Foundation.

\end{document}